\documentclass{article}
\usepackage{authblk}
\usepackage[utf8]{inputenc}
\usepackage{amsmath,amssymb,amsfonts,amsthm}
\usepackage{physics}
\usepackage{dsfont}
\usepackage{graphicx}
\usepackage{xcolor}
\usepackage{cancel}
\usepackage[numbers,sort&compress,square]{natbib}
\usepackage{color}
\usepackage{bbm}
\usepackage[linktocpage]{hyperref}
\usepackage{appendix}
\hypersetup{colorlinks=true,citecolor=blue,linkcolor=blue, urlcolor=blue, breaklinks=true}
\usepackage[eulergreek]{sansmath}
\usepackage{mathtools}
\usepackage{scalerel}

\usepackage{ulem}

\usepackage{bm} 
\usepackage{subfigure} 
\usepackage{environ}
\usepackage{url}
\usepackage{hyperref}
\usepackage[margin=1in]{geometry}

\usepackage{environ}
\NewEnviron{eqs}{%
\begin{equation}\begin{split}
    \BODY
\end{split}\end{equation}}

\newcommand{\ZZ}{{\mathbb Z}}
\newcommand{\Z}{{\mathbb Z}}
\newcommand{\R}{{\mathbb R}}

\newcommand{\eps}{\epsilon}

\graphicspath{{./Figures/}}

\begin{document}

\title{Onsiteability of Higher-Form Symmetries}

\author[1]{Yitao Feng}
\author[1]{Yu-An Chen}
\author[2]{Po-Shen Hsin}
\author[3, *]{Ryohei Kobayashi}

\affil[1]{International Center for Quantum Materials, School of Physics, Peking University, Beijing 100871, China}
\affil[2]{Department of Mathematics, King’s College London, Strand, London WC2R 2LS, UK}
\affil[3]{School of Natural Sciences, Institute for Advanced Study, Princeton, NJ 08540, USA}

\renewcommand*{\thefootnote}{*}
\footnotetext[1]{Contact author: ryok@ias.edu}

\renewcommand{\thefootnote}{\arabic{footnote}}
\date{\today}

\maketitle

\begin{abstract}
An internal symmetry in a lattice model is said to be onsiteable if it can be disentangled into an onsite action by introducing ancillas and conjugating with a finite-depth circuit. A standard lore holds that onsiteability is equivalent to being anomaly-free, which is indeed valid for finite 0-form symmetries in (1+1)D. 
However, for higher-form symmetries, these notions become inequivalent: a symmetry may be onsite while still anomalous.
In this work, we clarify the conditions for onsiteability of higher-form symmetries by proposing an equivalence between onsiteability and the possibility of \textit{higher} gauging.
For a finite 1-form symmetry in (2+1)D, we show that the symmetry is onsiteable if and only if its ’t Hooft anomaly satisfies a specific algebraic condition that ensures the symmetry can be 1-gauged.
We further demonstrate that onsiteable 1-form symmetry in (2+1)D can always be brought into transversal Pauli operators by ancillas and circuit conjugation. In generic dimensions, we derive necessary conditions for onsiteability using lattice 't Hooft anomaly of higher-form symmetry, and conjecture a general equivalence between onsiteability and possibility of higher gauging on lattices.
\end{abstract}

\tableofcontents

\section{Introduction}

Symmetry plays a central role in the study of quantum many-body systems and quantum field theory. In lattice models, symmetries not only constrain the possible dynamics but also determine the structure of low-energy excitations and quantum phases.
Among the various realizations, \textit{onsite} symmetries form a particularly important class.
Onsite symmetries correspond to the most physically natural implementation of internal symmetries in electronic and spin systems, where each local degree of freedom transforms independently under the symmetry action.

The notion of onsite symmetry also plays a key role in quantum information theory and fault-tolerant quantum computation. In that context, transversal single-qubit logical gates of quantum error-correcting codes form the analog of onsite operations: they act independently on physical qubits and therefore do not propagate local errors, ensuring fault tolerance.
The Eastin-Knill theorem~\cite{Eastin2009} establishes a no-go constraint on the set of logical operations that can be implemented transversally, motivating a systematic understanding of which symmetry operations can be represented in such strictly local, onsite fashion. From this perspective, a natural question arises: given a lattice realization of a global symmetry, is its action equivalent to an onsite action through conjugation by finite-depth circuits?

For ordinary 0-form symmetries in (1+1)D, there exists a well-established correspondence between the possibility of onsite realization and the absence of an ’t~Hooft anomaly that represents an obstruction to gauging a global symmetry~\cite{seifnashri2025disentangling}.
Concretely, a 0-form symmetry can be transformed into an onsite form by tensoring with ancillary Hilbert spaces and conjugating by a finite-depth circuit if and only if its ’t~Hooft anomaly is trivial on the lattice. Recent developments have further refined this correspondence through concise lattice formulations of the anomaly, providing a microscopic characterization of the ability to make a symmetry onsite~\cite{kawagoe2025anomaly, kapustin2025anomaly2d, kapustin2025higher, shirley2025QCA, tu2025anomaliesglobalsymmetrieslattice}.
The presence of 't Hooft anomalies also implies that the system cannot be realized in a short-range entangled (SRE) phase \cite{chen2010}.
For example, the Lieb-Schultz-Mattis theorem~\cite{LSM1961, oshikawa2, hastings2004, Kapustin2025LSM} and its generalizations~\cite{cheng2016set, cho2017anomaly, Kobayashi2019lsm, else2020lsm, prem2020lsm, seiberg2022lsm, Kapustin2025LSM, Liu2024spinS} enforce non-trivial constraints on the low-energy spectrum of lattice systems originating from mixed 't Hooft anomalies between spatial and internal symmetries.
Anomalies also enforce constraints on deconfinement in gauge theories~\cite{shimizu2018,hsin2020se,seifnashri2021sym} and lead to nontrivial edge states of symmetry-protected-topological (SPT) phases~\cite{Chen2011Twodimensional, chen2013SPT,ElseNayak2014,tiwari2018}.

For higher-form symmetries, however, the relation between being onsite and being anomaly-free becomes more subtle.
Higher-form symmetries act on extended objects such as lines or surfaces rather than on point-like degrees of freedom. Their ’t Hooft anomalies can likewise be defined and lead to rich dynamical consequences, much like those of ordinary symmetries. In particular, 't Hooft anomalies of higher-form symmetries enforce long-range entanglement of both pure and mixed states, which corresponds to refined dynamical constraints on the system \cite{Lessa:2024wcw, li2024anyon, lessa2025higher, zhou2025finiteT, Hsin:2025pqb}.

Recent advances have further enabled lattice formulations of higher-form anomalies~\cite{feng2025higherformanomalies, kobayashi2025generalizedstatistics, xue2024statistics, feng2025anyonic}.
Interestingly, a higher-form symmetry can remain anomalous in the field-theoretic sense yet still admit an onsite realization on the lattice. This observation raises a fundamental question: What is the correct general criterion for when a higher-form symmetry can be realized in an onsite manner?

To address this, we investigate \textit{onsiteability} of higher-form symmetry. A symmetry is said to be onsiteable if it can be transformed into a strictly onsite form by two types of operations:
\begin{enumerate}
    \item Tensoring with ancillary local degrees of freedom, each represented by a finite-dimensional onsite Hilbert space forming a representation of $G$.
    \item Conjugation by a local finite-depth quantum circuit acting on the enlarged Hilbert space of the original system and the ancilla.
\end{enumerate}

In this work, we clarify the onsiteability condition of higher-form symmetry, by proposing an equivalence between onsiteability and \textit{the possibility of higher gauging}~\cite{Roumpedakis2023higher}. Roughly speaking, higher gauging refers to gauging the higher-form symmetry within a submanifold of the whole space. This is regarded as a generalization of the correspondence between onsiteability and anomaly for higher-form symmetry.

For finite 1-form symmetries in (2+1)D, we prove this equivalence explicitly. The 't Hooft anomaly of the 1-form symmetry in (2+1)D on the lattice is characterized by an index $[\omega_4] \in H^4(B^2 G, U(1))$ \cite{feng2025higherformanomalies}. We show that the symmetry is onsiteable if and only if a cohomology operation $\Phi$ called a transgression of this anomaly index, $\Phi([\omega_4]) \in H^3(BG, U(1))$ is trivial.
A transgression $\Phi$ is a map of group cohomology described in the main text,
\begin{align}
    \Phi: H^{d+2}(B^{p+1}G,U(1))\to H^{d+1}(B^pG, U(1))~.
\end{align}
In a QFT, the transgression $\Phi([\omega])$ with an 't Hooft anomaly $\omega\in H^{d+2}(B^{p+1}G,U(1))$ characterizes the possibility of 1-gauging the symmetry; that is, the symmetry can be gauged within codimension-1 submanifolds of the system. When this condition is satisfied, the 1-form symmetry operators can be disentangled into an onsite form using ancillas and finite-depth circuits. Furthermore, we demonstrate that any onsiteable 1-form symmetry in (2+1)D can always be represented by transversal Pauli operators, showing that onsiteability implies a particularly simple realization familiar in Pauli stabilizer models such as toric codes. 
We remark that if we take $p=0$, this is the map that relates the (2+1)D Chern-Simons term for 1-form $G$ gauge fields and (1+1)D Wess-Zumino-Witten term for $G$-valued scalars as discussed in Ref.~\cite{Dijkgraaf:1989pz}.

To illustrate this criterion, consider the case of a $\mathbb{Z}_2$ 1-form symmetry in (2+1)D associated with the semion. 
The semion carries a topological spin of $1/4$, corresponding to a 1-form 't~Hooft anomaly characterized by~\cite{Tsui:2019ykk}
\begin{equation}
    \tfrac{1}{4}\bigl(B_2 \cup B_2 - B_2 \cup_1 \delta B_2\bigr) 
    \in H^4(B^2\mathbb{Z}_2, U(1)) = \mathbb{Z}_4,
\label{eq: 1/4 B2 B2}
\end{equation}
with a $\Z_2$ 2-form background $B_2$, which generates the $\mathbb{Z}_4$ classification of 1-form anomalies. 
Its transgression is the nontrivial cocycle
\begin{equation}
    \tfrac{1}{4} A_1 \cup \delta A_1 \in H^3(B\mathbb{Z}_2, U(1)) = \mathbb{Z}_2,
\label{eq:1/4 A1 dA1}
\end{equation}
with a $\Z_2$ 1-form background $A_1$, indicating that the corresponding 1-form symmetry is \textit{not} onsiteable.\footnote{In this simple example, the transgression map $\Phi: \mathbb{Z}_4 \to \mathbb{Z}_2$ acts by reduction modulo~2.} 
In contrast, the $\mathbb{Z}_2$ 1-form symmetry associated with the fermion, characterized by the cocycle $\tfrac{1}{2} B_2 \cup B_2$, 
has a trivial transgression in $H^3(B\mathbb{Z}_2, U(1))$, even though the symmetry remains anomalous. 
Hence, a fermionic 1-form symmetry is onsiteable. 
Indeed, in the $(2+1)$D $\mathbb{Z}_2$ toric code, the fermion excitation is onsite: its symmetry operator is realized as a product of Pauli $X$ and $Z$ operators along a closed string.

We then extend the above result to higher-form symmetry in generic dimensions. In $(d+1)$ spacetime dimensions higher than (1+1)D, onsiteability of 0-form symmetry is not only obstructed by the continuum QFT 't Hooft anomalies $H^{d+2}(BG,U(1))$, but additional ``lattice anomaly'' indices valued in $H^{d+2-q}(BG, \mathrm{QCA}_{q-1})$~\cite{shirley2025QCA, tu2025anomaliesglobalsymmetrieslattice}. Here, $\mathrm{QCA}_{q-1}$ denotes equivalence classes of quantum cellular automata (QCAs) in $(q-1)$ spatial dimensions~\cite{GNVW2012, Haah:2018jdf, Haah2021CliffordQCA, Shirley2022ThreeQCA, fidkowski2024qca, sun2025clifford}. These indices generalize the continuum ’t Hooft anomaly by capturing obstructions to onsiteability or gauging that exist only at the microscopic lattice models. 

We discover such ``lattice anomaly'' indices of higher-form symmetries. In particular, for a finite 1-form symmetry in (3+1)D, we explicitly define an index in $H^3(B^2 G, \mathrm{QCA}_{1})$, whose transgression $\Phi(\omega_3) \in H^2(BG, \mathrm{QCA}_{1})$ diagnoses the obstruction to onsiteability. This index $\Phi(\omega_3)$ is thought of as an obstruction to 1-gauging the symmetry on the lattice, implying that the correspondence between onsiteability and higher gauging is valid at the level of these additional ``lattice anomaly'' indices beyond the QFT anomalies.

We conjecture a general criterion for the onsiteability of finite higher-form symmetries, formulated through higher gauging in lattice systems:
\begin{quote}
Consider a finite $p$-form symmetry group $G$ in $(d+1)$ spacetime dimensions. 
We define a sequence of ``lattice anomaly'' indices
$[\omega_{d+2-q}] \in H^{d+2-q}(B^{p+1}G, \mathrm{QCA}_{q-1})$
for $q = 0, 1, 2, \ldots, d+1$.\footnote{Here, $\mathrm{QCA}_{-1} := U(1)$ reproduces the conventional 't~Hooft anomaly when $q = 0$.} 
Their successive transgressions,
$\Phi^p([\omega_{d+2-q}]) \in H^{d+2-p-q}(BG, \mathrm{QCA}_{q-1})$,
jointly characterize the obstructions to performing $p$-gauging of the symmetry on the lattice. 
A $p$-form symmetry is onsiteable if and only if all such obstructions vanish:
\begin{equation}
    \Phi^p([\omega_{d+2-q}]) = 0 
    \in H^{d+2-p-q}(BG, \mathrm{QCA}_{q-1}), 
    \qquad  \forall~ 0 \le q \le d+1~.
\end{equation}
\end{quote}

This paper is organized as follows. In Sec.~\ref {sec:2+1D} we define the transgression of group cohomology, and describe the onsiteability condition of 1-form symmetry in (2+1)D. In Sec.~\ref {sec:higherdim} we extend the onsiteability criteria to higher-form symmetry in generic dimensions, including the discussions of lattice anomaly indices.

\section{Onsiteability of 1-form symmetries in (2+1)D}
\label{sec:2+1D}

\subsection{Transgression of group cohomology}

We now define the transgression map, which plays a central role in diagnosing the onsiteability of higher-form symmetries:
\begin{align}
    \Phi: H^{d+2}(B^{p+1}G, U(1)) \to H^{d+1}(B^pG, U(1))~.
\end{align}
Let $[\omega_{d+2}] \in H^{d+2}(B^{p+1}G, U(1))$ denote a cohomology class represented by a cocycle functional $\omega_{d+2}[B_{p+1}]$ of a $(p+1)$-form background field $B_{p+1}$ associated with the symmetry group $G$. 
When the background is exact, $B_{p+1} = \delta B_p$, the cocycle satisfies 
$\omega_{d+2}[\delta B_p] = \delta \omega_{d+1}[B_p]$ 
for some $(d+1)$-cochain $\omega_{d+1}$. 
Evaluating $\omega_{d+1}[B_p]$ on closed configurations with $\delta B_p = 0$ then defines a class in $H^{d+1}(B^pG, U(1))$. 
We therefore define the transgression as $\Phi([\omega_{d+2}]) := [\omega_{d+1}]$. 
This construction has appeared previously in, for example, Refs.~\cite{Kaidi2022KW, Choi2023triality, Roumpedakis2023higher}, 
in the context of non-invertible symmetries in (3+1)D $(d=2, p=1)$.

\paragraph{Transgression and higher gauging}

The cohomology $[\omega_{d+2}]\in H^{d+2}(B^{p+1}G,U(1))$ describes an 't Hooft anomaly of $p$-form $G$ symmetry in $(d+1)$ spacetime dimensions, which signals an obstruction to gauging the $p$-form symmetry. Meanwhile, its transgression $\Phi(\omega_{d+2})$ describes an obstruction to ``1-gauging'' the symmetry \cite{Roumpedakis2023higher}; consider a codimension-1 submanifold of the whole $(d+1)$D spacetime, and let us attempt to gauge the symmetry within this submanifold. Then the symmetry operators generate $(p-1)$-form symmetry within the submanifold, and its obstruction to gauging is captured by $[\Phi(\omega_{d+2})]\in H^{d+1}(B^pG, U(1))$.

To see this, let us consider a $(d+1)$D QFT $\mathcal{T}_{d+1}$ with $p$-form $G$ symmetry that has an 't Hooft anomaly $[\omega_{d+2}]\in H^{d+2}(B^{p+1}G, U(1))$.
We put $\mathcal{T}_{d+1}$ on a $(d+1)$-manifold $M_{d+1}$ with a boundary $N_d=\partial M_{d+1}$; we take the boundary condition such that the $p$-form symmetry operators in the bulk again defines a nontrivial topological operator at the boundary, which generates $(p-1)$-form $G$ symmetry at the $d$-dimensional boundary. Let us denote the background $G$ gauge fields at the boundary and bulk by $B_p$, $B_{p+1}$ respectively. They are subject to the boundary condition $B_{p+1}|=\delta B_p$ at the boundary $N_d$.

The inflow of 't Hooft anomaly in the bulk is represented by a $(d+2)$D response $\omega_{d+2}$. On the boundary $M_{d+1}$, the inflow response action $\omega_{d+1}$ in $(d+1)$D satisfies $\omega_{d+2}[\delta B_{p}]| = \delta\omega_{d+1}[B_{p}]$ due to the gauge invariance of the bulk-boundary response action. Now let us turn off the gauge field in the bulk, $B_{p+1}=0$. Then, the above boundary condition $\omega_{d+2}[\delta B_{p}]| = \delta\omega_{d+1}[B_{p}]$ implies that $\omega_{d+1}$ is a transgression of $\omega_{d+2}$. Therefore $[\omega_{d+1}]=\Phi([\omega_{d+2}])$ describes the 't Hooft anomaly of $(p-1)$-form symmetry generated by the topological operators restricted within the $d$-dimensional boundary. This implies that $\Phi([\omega_{d+2}])\in H^{d+1}(B^pG,U(1))$ describes an obstruction to 1-gauging the symmetry at a $d$-manifold $N_d$.

\subsection{Review: Anomaly index and onsite 1-form symmetry}
\label{sec:reviewof1formanomaly}

Let us focus on 1-form $G$ symmetry in (2+1)D. 
We introduce 1-form symmetry in a 2d tensor product Hilbert space in most generic setup, and define the anomaly index $[\omega_4]\in H^4(B^2G, U(1))$. This is a review of Ref.~\cite{feng2025higherformanomalies}.

\begin{figure}[htb]
\centering
\includegraphics[width=0.7\textwidth]{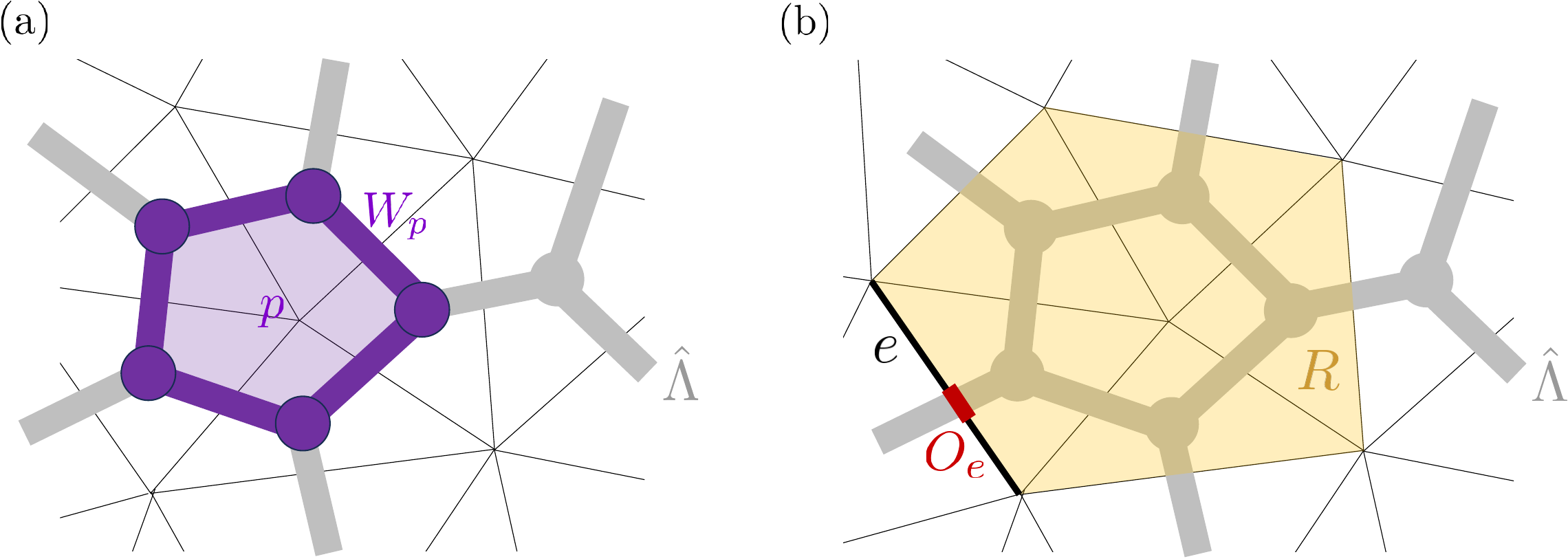}
\caption{(a): Symmetry operators are defined on a mesoscopic dual lattice $\hat\Lambda$. (b): The local operator $O_e$ is supported at the intersection between edges of $\Lambda$ and $\partial R$. }
\label{fig:duallattice}
\end{figure}

We begin with a two-dimensional spatial lattice endowed with a tensor product Hilbert space, upon which one seeks to implement a finite 1-form $G$ symmetry generated by finite-depth circuits localized near codimension-1 regions. We introduce a “mesoscopic” triangulation $\Lambda$ in the space, whose edges are chosen large compared to both the circuit depth and the microscopic locality length. The symmetry circuits will are supported within a thin strip along the mesoscopic dual lattice $\hat\Lambda$, see Fig.~\ref{fig:duallattice}.
One associates to each plaquette \(p\) a Gauss law operator \(W^{(g)}_p\), labeled by group elements \(g\in G\). A single plaquette $p$ of $\hat\Lambda$ is dual to a vertex of $\Lambda$.
These operators are small loops of symmetry generators, and obey the group algebra on each plaquette:
\begin{equation}
W^{(g)}_p \,W^{(g')}_p \;=\; W^{(gg')}_p \,.
\label{eq:gausslaw1}
\end{equation}
They mutually commute on distinct plaquettes,
\begin{equation}
\bigl[\,W^{(g)}_p,\; W^{(g')}_{p'}\bigr] = 1\,,
\end{equation}
with a group commutator $[U,V]:= U^{-1}V^{-1}UV$, and a global constraint is imposed that the product over all plaquettes yields the identity:
\begin{equation}
\prod_p W^{(j)}_p = 1 \quad \forall\; j \in G\,.
\label{eq:globalconstraint}
\end{equation}
This ensures that the symmetry operators defined on closed loops are topological. The above conditions complete the definition of 1-form $G$ symmetry in (2+1)D.

One then considers general 0-cochains \(\epsilon \in C^0(\Lambda, G)\) (decomposed into components \(\epsilon = \oplus_j \epsilon_j\) for each cyclic factor of \(G\)). A global symmetry operator is built as
\begin{equation}
U(\epsilon) = \prod_p \bigl(W_p\bigr)^{\epsilon(p)}\,,
\label{eq:def of 1form symmetry operators}
\end{equation}
which satisfies the condition
\begin{equation}
U(\epsilon + dg) = U(\epsilon)~,
\end{equation}
where $dg\in C^0(\Lambda,G)$ is a constant (global) cochain with the constant value \(g\in G\) at each vertex.  

Next, one restricts \(U(\epsilon)\) to a disk region $R$ of $\Lambda$. Denoting this restricted unitary by \(U_R(\epsilon)\), we choose a concrete truncation of the Gauss law operators along the boundary \(\partial R\), denoted \(W_{p;R}\), such that 
\begin{equation}
U_R(\epsilon) = \Bigl(\prod_{p\in \partial R} W_{p;R}^{\epsilon(p)}\Bigr)\;\Bigl(\prod_{p \in \mathrm{Int}(R)} W_p^{\epsilon(p)}\Bigr).
\end{equation}
Because the truncated operators \(W_{p;R}\) need not commute, one must impose ordering of operators to define their product. The anomaly index is independent of this ordering.  
One then defines a reduced operator  
\begin{equation}
\Omega(\epsilon_{01},\epsilon_{12}, g_{012}) \;=\; U_R(\epsilon_{01})\, U_R(\epsilon_{12})\, U_R(\epsilon_{01} + \epsilon_{12} - dg_{012})^{-1}\,,
\end{equation}
with $\epsilon_{01},\epsilon_{12} \in C^0(\Lambda, G)$, $g_{012}\in G$.
This can be further decomposed into separate local factors along edges \(e\) of the boundary:
\begin{equation}
\Omega(\epsilon_{01},\epsilon_{12}, g_{012}) = \prod_{e\in \partial R} \, O_e(\epsilon_{01}, \epsilon_{12}, g_{012})\,.
\end{equation}
See Fig.~\ref{fig:duallattice} (b).
From the associativity property of the unitaries one obtains a “2-cocycle equation” satisfied by $\Omega$:
\begin{equation}
\Omega(\epsilon_{01},\epsilon_{12})\Omega(\epsilon_{02},\epsilon_{23})
    =~^{\epsilon_{01}}\Omega(\eps_{12},\eps_{23}) \Omega(\eps_{01},\eps_{13})~,
\end{equation}
where $^{\epsilon}O = U_R(\epsilon) O U_R(\epsilon)^{-1}$.
Now we can introduce a local functional defined on each edge $e$ of $\partial R$,
\begin{equation}
F_e(\epsilon_{01},\epsilon_{12},\epsilon_{23}, \{g\}) \in \R/\Z\,,
\end{equation}
defined via
\begin{equation}
e^{2\pi i F_e(\epsilon_{01},\eps_{12},\eps_{23},\{g\})}:= O_e(\epsilon_{01},\epsilon_{12})O_e(\epsilon_{02},\epsilon_{23})\left(^{\epsilon_{01}}O_e(\eps_{12},\eps_{23})O_e(\eps_{01},\eps_{13})\right)^{-1}~.
\end{equation}
These local \(F_e\) form a 1-cocycle on \(\partial R\), whose integral vanishes:
\begin{equation}
\sum_{e \in \partial R} F_e(\ldots) = 0 \quad (\bmod\, 1)\,,
\end{equation}
hence \(F = dA\) is exact for some 0-cochain \(A\) supported on vertices along \(\partial R\).  Restricting to an interval \(I \subset \partial R\) and comparing products along left and right sub-intervals yields
\begin{equation}
\Omega_I(\epsilon_{01},\epsilon_{12})\Omega_I(\epsilon_{02},\epsilon_{23}) (^{\epsilon_{01}}\Omega_I(\epsilon_{12},\epsilon_{23})\Omega_I(\epsilon_{01},\epsilon_{13}))^{-1} = e^{2\pi i \int_{I}F_e(\epsilon_{01},\eps_{12},\eps_{23},\{g\})} = e^{2\pi i (A_l-A_r)}~,
\end{equation}
where \({l}\) and \(r\) label the interval endpoints.  

The “left endpoint” functional \(A_{l}\) then satisfies a 3-cocycle condition following from that satisfied by $F_e$:
\begin{equation}
F_e(\epsilon_{01},\epsilon_{12},\epsilon_{23}) + F_e(\epsilon_{01},\epsilon_{13},\epsilon_{34}) + F_e(\epsilon_{12},\epsilon_{23},\epsilon_{34}) = F_e(\epsilon_{02},\epsilon_{23},\epsilon_{34}) + F_e(\epsilon_{01},\epsilon_{12},\epsilon_{24})  \quad \text{mod $1$,}
\end{equation}
which in turn implies \(A_{l}\) obeys a related 3-cocycle relation.  From this one constructs a 4-cochain (the anomaly index)
\begin{equation}
\begin{split}
\omega_{4;l}(\epsilon_{01},\epsilon_{12},\epsilon_{23},\epsilon_{34},\{g\}) := \delta A_l &= A_l(\epsilon_{01},\epsilon_{12},\epsilon_{23}) + A_l(\epsilon_{01},\epsilon_{13},\epsilon_{34}) + A_l(\epsilon_{12},\epsilon_{23},\epsilon_{34})  \\
&\quad 
- A_l(\epsilon_{02},\epsilon_{23},\epsilon_{34}) - A_l(\epsilon_{01},\epsilon_{12},\epsilon_{24})~.
\label{eq:def of omega4}
\end{split}
\end{equation}
One can show that \(\omega_{4;l}\) is independent of the cochains \(\epsilon\), which depends only on the group labels \(g_{ijk}\). $\omega_{4;l}$ is also independent of the vertex $l$, therefore one can simply write it as $\omega_4$. The index further satisfies the 4-cocycle condition, and possible ambiguities shift $\omega_4$ by a coboundary. Therefore the index is valued in $H^4(B^2 G, U(1))$ and defines a cohomology class
\begin{equation}
[\omega_4] \in H^4(B^2 G, U(1))\,.
\end{equation}

\paragraph{Onsite 1-form symmetry}
Onsite 1-form $G$ symmetry means that with a suitable tensor product decomposition into local onsite Hilbert spaces, 
each Gauss law operator $W_p^{(g)}$ is expressed as a product of operators that act on onsite Hilbert spaces,
\begin{align}
    W_p^{(g)} = \bigotimes_{j\in \partial p} U^{(g)}_j~,
\end{align}
where $j$ labels the onsite Hilbert space $\mathcal{H}_j$ supported within a thin strip along $\partial p$.

\subsection{Onsiteability condition}
\label{sec:onsiteability 2+1D}

We establish the following statement:
\begin{quote}
A 1-form $G$ symmetry in $(2+1)$D is onsiteable if and only if the transgression
$\Phi: H^4(B^2 G,U(1)) \rightarrow H^3(BG,U(1))$
of its anomaly class $[\omega_4]\in H^4(B^2 G,U(1))$ vanishes in cohomology:
\[
[\Phi(\omega_4)]=0 \in H^3(BG,U(1)).
\]
Equivalently, the symmetry is onsiteable precisely when it is 1-gaugeable.
\end{quote}

\paragraph{``Only if'' part} Let us consider a 1-form symmetry operator, extended along a thin strip of a 1-cycle $\hat{\gamma}$ of $\hat{\Lambda}$. This operator is interpreted as a 0-form $G$ symmetry acting on a 1d tensor product Hilbert space along the 1-cycle $\hat{\gamma}$. Therefore, by using the standard method by Else and Nayak \cite{else2014} (see Appendix \ref{app:elsenayak} for a review), one can define an index $[\omega_3]\in H^3(BG,U(1))$. According to Ref.~\cite{seifnashri2025disentangling}, this line operator is onsiteable if and only if $[\omega_3] = 0$.

We now show that if the 1-form symmetry is onsiteable, the transgression of its bulk anomaly must vanish:
\begin{enumerate}
    \item \textbf{Global onsiteability.}  
    Suppose the 1-form symmetry is onsiteable. 
    Then there exists a finite-depth quantum circuit (FDQC), together with local ancillas on the 2d lattice, that transforms the symmetry into an onsite operator. 
    The bulk anomaly class $[\omega_4]\in H^4(B^2G,U(1))$ is invariant under such FDQCs (and under adding trivial ancillas), so its transgression $[\Phi(\omega_4)]$ is likewise unchanged.

    \item \textbf{Restriction to a line.}  
    Since the symmetry becomes onsite on the 2d plane, its restriction to any 1d line or strip is also onsite. 
    Hence the corresponding Else--Nayak index vanishes:
    \[
        [\omega_3] = 0 \in H^3(BG,U(1)).
    \]

    \item \textbf{Relation to transgression.}  
    As shown below, the Else--Nayak index equals the transgression of the bulk anomaly,
    \[
        [\omega_3] = \Phi([\omega_4])~,
    \]
    establishing that
    \[
        \text{onsiteable 1-form symmetry} \;\Rightarrow\; [\Phi(\omega_4)] = 0~.
    \]
\end{enumerate}

\begin{figure}[thb]
\centering
\includegraphics[width=0.45\textwidth]{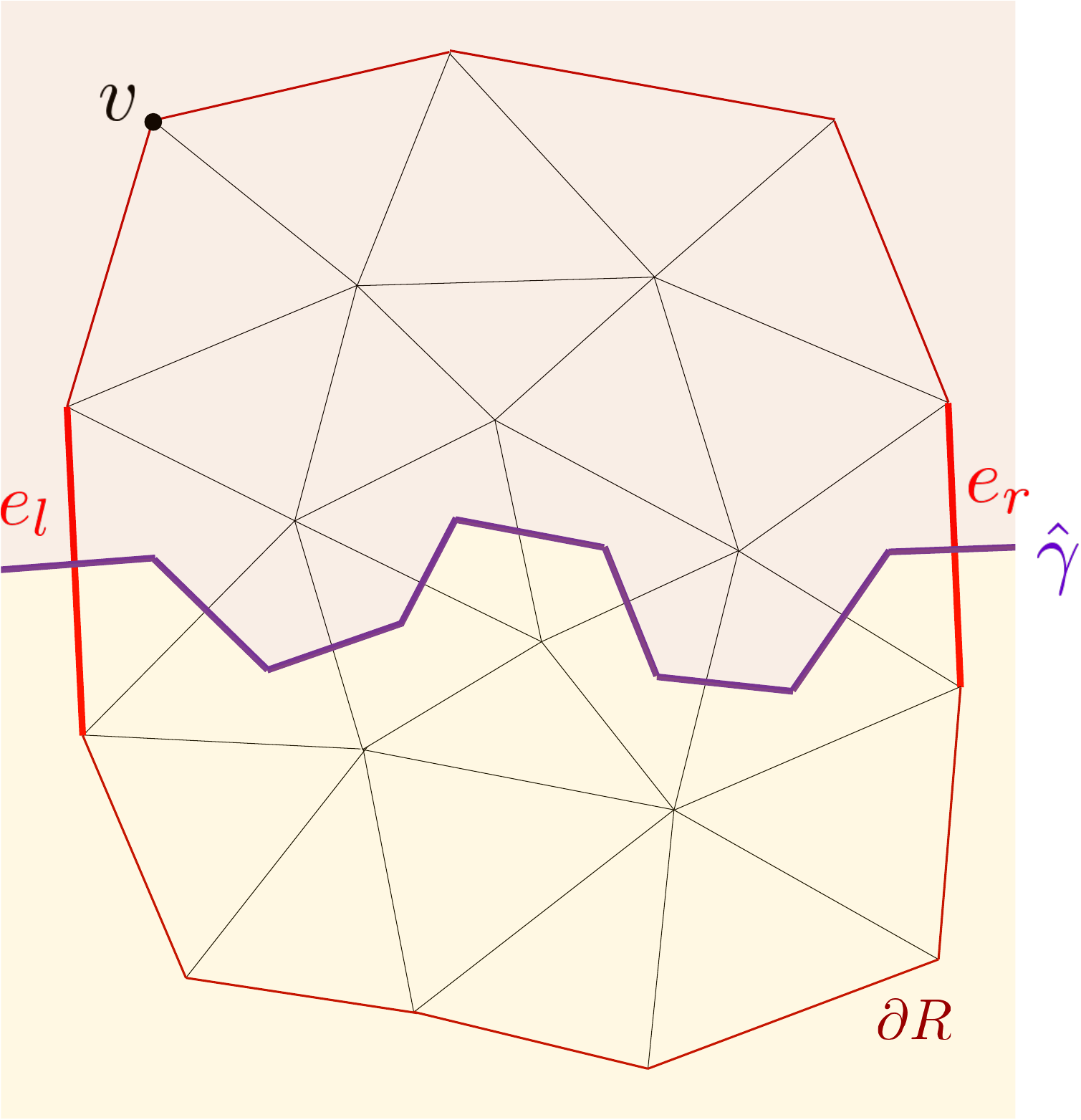}
\caption{A cycle $\hat{\gamma}$ of the dual lattice separates the disk region $R$ into half, up ($u$) and down ($d$) region.}
\label{fig:bipartition}
\end{figure}

The remaining task is to show that the above $[\omega_3]\in H^3(BG,U(1))$ is a transgression of the anomaly index $[\omega_4]\in H^4(B^2G,U(1))$.
Let us first consider the 1-cycle of the dual lattice $\hat\gamma$ which cuts $R$ into a bipartition. See Fig.~\ref{fig:bipartition}. 
The region $R$ is separated into half disks $R_u,R_d$, and $\partial R$ into intervals $(\partial R)_u, (\partial R)_d$. 

Then, for a given constant $x$, we define a 1-cochain $\iota_*x$ on $\Lambda$ by
\begin{align}
    \iota_*x(v) = 
    \begin{cases}
        x & \text{$v\in \Lambda_u$} \\
        0 & \text{$v\in \Lambda_d$}~.
    \end{cases}
\end{align}
In particular when $h\in G$, taking the symmetry operator $U(\iota_*h)$ gives a line operator along the 1-cycle $\hat{\gamma}$ generating the $G$ symmetry.

The 1-cycle $\hat{\gamma}$ intersects with $\partial R$ at a pair of edges of $\Lambda$, $e_l$ and $e_r$. By regarding the operator $U(\iota_*h)$ as a 0-form $G$ symmetry operator in 1d, $F_{e_l}(\iota_*h_{01},\iota_*h_{12},\iota_*h_{23})$ with $h_{01},h_{12},h_{23}\in G$ is the Else-Nayak index of the 1d 0-form symmetry~\cite{else2014} ($\delta h_{ijk} := h_{ij}+ h_{jk} - h_{ik}=0$ is understood);
\begin{align}
F_{e_l}(\iota_*h_{01},\iota_*h_{12},\iota_*h_{23}) = \omega_3(h_{01},h_{12}, h_{23})~,
\end{align}
which directly follows from its construction (see Appendix \ref{app:elsenayak} for a review of the Else-Nayak index). 
Due to the same reasoning, 
\begin{align}
F_{e_r}(\iota_*h_{01},\iota_*h_{12},\iota_*h_{23}) = -\omega_3(h_{01},h_{12}, h_{23})~.
\end{align}
Note that $F(\iota_*h_{01},\iota_*h_{12},\iota_*h_{23})=0$ except for the above two edges $e_l,e_r$. Using the operation $\iota_*$ introduced above, $F$ is simply expressed as
\begin{align}
    F(\iota_*h_{01},\iota_*h_{12},\iota_*h_{23}) = d(\iota_*\omega_3)(h_{01},h_{12},h_{23})~.
\end{align}
Since $F=dA$, we get
\begin{align}
    A(\iota_*h_{01},\iota_*h_{12},\iota_*h_{23}) = \iota_*\omega_3(h_{01},h_{12},h_{23})~.
\end{align}
Let $v$ be a vertex of $(\partial R)_u$. Then 
\begin{align}
    A_v(\iota_*h_{01},\iota_*h_{12},\iota_*h_{23}) = \omega_3(h_{01},h_{12},h_{23})~.
    \label{Av=omega3}
\end{align}
Meanwhile, for generic 0-cochains $\epsilon_{ij}\in C^0(\Lambda,G)$, due to \eqref{eq:def of omega4} we have
\begin{align}
    \delta A_v(\epsilon_{01},\epsilon_{12},\epsilon_{23},\epsilon_{34}) = \omega_4~.
\end{align}
Let us write the values $\epsilon_{ij}(v)= h_{ij}\in G$. 
In general $h_{ij}$ is not closed under the coboundary $\delta$;
$g_{ijk} = (\delta h)_{ijk}$. Also, $A_v$ depends on $\epsilon_{ij}$ through their values at $v$, therefore one can write
\begin{align}
    \delta A_v(\{h\}) = \omega_4(\{\delta h\})~.
\end{align}
Since $\omega_4\in Z^4(B^2G,U(1))$, $\omega(\delta h)$ is expressed as a coboundary using some 3-cochain $\chi$,
\begin{align}
    \omega_4(\{\delta h\}) = \delta \chi(\{h\})~.
    \label{omega4 = dchi}
\end{align}
This implies that 
\begin{align}
    A_v(\{h\}) = \chi(\{h\})~.
\end{align}
Now let us take $\{h\}$ such that $\delta h=0$. Due to \eqref{Av=omega3}, by writing $A_v$ as a functional of $\epsilon_{ij}(v)$ we have
\begin{align}
    A_v(\{h\}) = \omega_3(\{h\}) \quad \text{when $\delta h=0$~.}
\end{align}
By the above two equations, we get
\begin{align}
    \omega_3(\{h\})=\chi(\{h\}) \quad \text{when $\delta h=0$~.}
    \label{omega3 = chi}
\end{align}
Eqs.~\eqref{omega4 = dchi}, \eqref{omega3 = chi} together imply that $\Phi(\omega_4)=\omega_3$. This shows that $\Phi(\omega_4)$ must vanish for onsiteability.

\begin{figure}[thb]
\centering
\includegraphics[width=\textwidth]{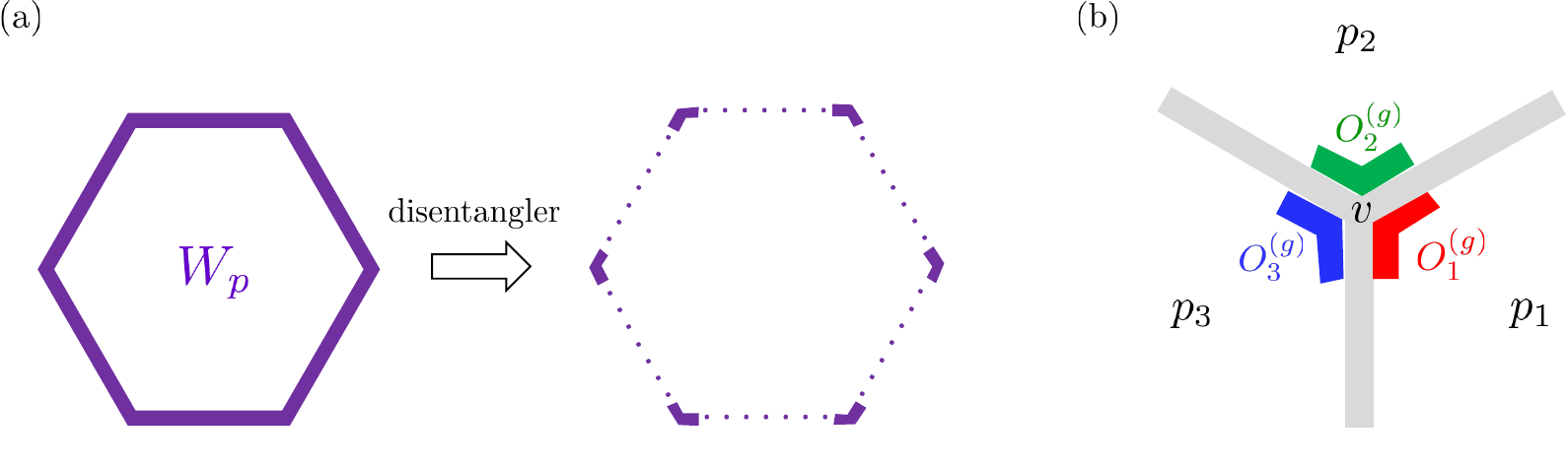}
\caption{(a) By introducing a 1d ancilla and disentangler on each mesoscopic edge, the Gauss law operator $W_p$ is brought into onsite form.
(b) After bringing into onsite form, each vertex has three local operators $O_j^{(g)}$ ($j=1,2,3$) that are local symmetry operators at the corners of each plaquette.}
\label{fig:disentangler}
\end{figure}

\paragraph{``If'' part.}  
The converse follows by explicitly disentangling the 1-form symmetry operators using the 1d circuits constructed in Ref.~\cite{seifnashri2025disentangling}. 
Along each edge $e$ of the mesoscopic dual lattice $\hat{\Lambda}$, we introduce an ancillary 1d tensor-product Hilbert space $\mathcal{H}'_e$ and a local disentangler $V_e$ acting on $\mathcal{H}_e \otimes \mathcal{H}'_e$, 
where $\mathcal{H}_e$ is the original 1d Hilbert space on which the 1-form symmetry acts. 
Since the 1d symmetry operator along $e$ has a trivial index $[\omega_3]=0$, one can construct a disentangler that transforms each Gauss-law operator $W^{(g)}_p$ ($g\in G$) into an onsite form:
\begin{align}
    V \!\left(W^{(g)}_p \otimes \!\bigotimes_j \mathcal{X}'^{(g)}_j\!\right)\! V^\dagger 
    = \!\left(\bigotimes_{v\subset \partial p} O^{(g)}_v\!\right) 
      \otimes \!\left(\bigotimes_j \mathcal{X}'^{(g)}_j\!\right),
    \label{eq:onsite1form}
\end{align}
where $V = \bigotimes_e V_e$. 
Here, $v$ labels the vertices of $\hat{\Lambda}$ along $\partial p$, and $O^{(g)}_v$ denotes local operators supported near each vertex $v$. 
The index $j$ runs over the remaining onsite Hilbert spaces along $\partial p$ that avoid the vertex loci. 
The operators $\mathcal{X}'^{(g)}_j$ are onsite symmetry operators acting on the ancilla Hilbert spaces $\mathcal{H}'_e$ (see Fig.~\ref{fig:disentangler}(a)).
By treating the collection of vertices $\{v\}$ as the new onsite degrees of freedom, the resulting symmetry becomes a manifestly onsite 1-form $G$ symmetry. 
This completes the proof.

\subsection{Onsite 1-form symmetries are realized by Pauli operators}
\label{subsec:pauli}

We now show that if a $G$ 1-form symmetry in (2+1)D is onsiteable, it can be further transformed—using ancillas and disentanglers—into transversal Pauli operators. 
For simplicity, we establish this for $G=\mathbb{Z}_N$; the generalization to arbitrary finite Abelian groups is straightforward. 
When $G=\mathbb{Z}_N$, the symmetry operators can be written as products of $\mathbb{Z}_N$ Pauli operators $\{X,Z\}$, which are $N\times N$ matrices acting on $\mathbb{Z}_N$ qudits and satisfying the Pauli algebra
\begin{align}
    X^N = Z^N = 1~, \quad XZ = e^{\frac{2\pi i}{N}} ZX~.
\end{align}
In the onsite form of the symmetry operators in Eq.~\eqref{eq:onsite1form}, each ancillary onsite Hilbert space can be chosen to be $N$-dimensional, 
and each operator $\mathcal{X}'^{(g)}_j$ with $g=1\in\mathbb{Z}_N=\{0, 1, \cdots N-1\}$ can be identified with the $\mathbb{Z}_N$ Pauli $X$ operator~\cite{seifnashri2025disentangling},
\begin{align}
    \mathcal{X}'^{(1)} = X~.
\end{align}
The remaining task is to show that the local operators $O^{(g)}_v$ near a vertex $v$ of $\hat{\Lambda}$ are transformed into Pauli operators.

Consider a single vertex $v$ of $\hat{\Lambda}$ surrounded by three plaquettes $p_1, p_2, p_3$. 
By acting the disentangler $V$ on the Gauss-law operators, each Gauss law on plaquette $p_j$ becomes a local operator $O^{(g)}_j$ at $v$ (see Fig.~\ref{fig:disentangler}(b)). 
These operators satisfy the $\mathbb{Z}_N$ algebra
\begin{align}
    \left(O^{(1)}_j\right)^N = 1~.
    \label{eq:O^N=1}
\end{align}
Moreover, since the Gauss-law operators $W_p^{(g)}$ commute, the commutators between $O^{(1)}_j$ and $O^{(1)}_k$ for $g=1\in\mathbb{Z}_N$ yield phase factors,
\begin{align}
    [O^{(1)}_j, O^{(1)}_k] = e^{\frac{2\pi i}{N} n_{j,k}}~,
    \label{eq:commutation of O}
\end{align}
where $n_{j,k}\in\mathbb{Z}_N$ and $j,k\in\{1,2,3\}$. 
From now on, we denote $O^{(1)}_j$ simply as $O_j$.

The algebra in Eqs.~\eqref{eq:O^N=1} and~\eqref{eq:commutation of O} can be realized by products of $\mathbb{Z}_N$ Pauli operators. 
Consider an $N^3$-dimensional Hilbert space, corresponding to three $\mathbb{Z}_N$ qudits labeled R, G, and B. 
We introduce the operators
\begin{align}
    O'_1 = X_\text{G} (Z_\text{B})^{n_{2,1}}~, \quad 
    O'_2 = X_\text{B} (Z_\text{R})^{n_{3,2}}~, \quad 
    O'_3 = X_\text{R} (Z_\text{G})^{n_{1,3}}~.
\end{align}
These operators obey the same algebra as in Eqs.~\eqref{eq:O^N=1} and~\eqref{eq:commutation of O}.

Now, we construct explicit ancillas and disentanglers that transform the operators $O_j$ into Pauli operators. 
At each vertex $v$, we introduce an $N^6$-dimensional ancilla Hilbert space, equivalent to six $\mathbb{Z}_N$ qudits labeled 
R, G, B, $\overline{\text{R}}$, $\overline{\text{G}}$, $\overline{\text{B}}$. 
The total Hilbert space at a vertex $v$ is
\begin{equation}
    \mathcal{H}_v \otimes \mathcal{H}'_v \otimes \mathcal{H}''_v 
    = (\text{original}) \otimes (\text{R,G,B}) \otimes (\overline{\text{R}}, \overline{\text{G}}, \overline{\text{B}})~.
\end{equation}
A local disentangler $V_v$ acting at vertex $v$ is constructed such that
\begin{align}
\begin{split}
    V_v\!\left( O_1 \otimes O_1'^\dagger \otimes X_{\overline{\text{R}}} \right)\! V_v^\dagger &= X_{\overline{\text{R}}}~, \\
    V_v\!\left( O_2 \otimes O_2'^\dagger \otimes X_{\overline{\text{G}}} \right)\! V_v^\dagger &= X_{\overline{\text{G}}}~, \\
    V_v\!\left( O_3 \otimes O_3'^\dagger \otimes X_{\overline{\text{B}}} \right)\! V_v^\dagger &= X_{\overline{\text{B}}}~.
\end{split}
\end{align}
Because the operators $U_j = O_j \otimes O_j'^\dagger \otimes I$ commute, the disentangler $V_v$ can be written as a product of Controlled-$U_j$ gates:
\begin{align}
    V_v = (CU)_1 (CU)_2 (CU)_3~,
\end{align}
where $(CU)_1$ is defined as
\begin{align}
    (CU)_1\!\left(\ket{\psi} \otimes \ket{l}_{\overline{\text{R}}}\right) 
    = (O_1 \otimes O_1'^\dagger)^l \ket{\psi} \otimes \ket{l}_{\overline{\text{R}}}~,
\end{align}
with $\ket{\psi}$ denoting the state of all qudits at $v$ except for $\overline{\text{R}}$, 
and $\ket{l}_{\overline{\text{R}}}$ the Pauli-$Z$ eigenstate of qudit $\overline{\text{R}}$:
\begin{align}
    Z_{\overline{\text{R}}} \ket{l}_{\overline{\text{R}}} = e^{\frac{2\pi i}{N}l} \ket{l}_{\overline{\text{R}}}~, \quad l \in \mathbb{Z}_N~.
\end{align}
The definitions of $(CU)_2$ and $(CU)_3$ follow analogously.

At each vertex $v$, the Gauss law operators tensored with onsite ancillary operators, 
$O_1 \otimes X_{\overline{\text{R}}}$, $O_2 \otimes X_{\overline{\text{G}}}$, and $O_3 \otimes X_{\overline{\text{B}}}$, 
are conjugated by $V_v$ and thereby transformed into the Pauli operators 
$O'_1 \otimes X_{\overline{\text{R}}}$, $O'_2 \otimes X_{\overline{\text{G}}}$, and $O'_3 \otimes X_{\overline{\text{B}}}$.
Combining this with Eq.~\eqref{eq:onsite1form}, the total disentangler
\begin{align}
    V = \bigotimes_e V_e \bigotimes_v V_v
\end{align}
maps every Gauss law operator, together with its onsite ancillas, into a product of onsite Pauli operators.

\subsection{Example: semion is not onsiteable}
\label{subsec:semion}

While the anomaly of a $\mathbb{Z}_N$ 1-form symmetry is characterized by the index 
$[\omega_4]\in H^4(B^2\mathbb{Z}_N,U(1))$, it can equivalently be described by the 
T-junction invariant~\cite{Levin2003Fermions, Liujun2024Symmetriesanomalies, feng2025higherformanomalies}. 
When the Gauss-law operators take the onsite form~\eqref{eq:onsite1form}, the T-junction invariant can be expressed as a product of $O$ operators at a single vertex $v$:
\begin{align}
    e^{2\pi i\Theta} 
    = O_1^{\dagger} O_2 O_3^{\dagger} O_1 O_2^{\dagger} O_3 
    = [O_3,O_2][O_2,O_1][O_1,O_3]~.
\end{align}
This invariant determines the spin $\Theta$ of the anyon that generates the 1-form symmetry operator. 
Since the operators $O_j$ satisfy the algebra~\eqref{eq:commutation of O}, 
an onsiteable $\mathbb{Z}_N$ symmetry operator must have spin
\begin{align}
    \Theta = \frac{n}{N}~, \qquad n \in \mathbb{Z}_N~.
\end{align}
For example, when $N=2$, an onsiteable 1-form $\mathbb{Z}_2$ symmetry can only be generated by anyons 
with spin $\Theta = 0$ or $\tfrac{1}{2}$. 
Therefore, the semion, with a $\mathbb{Z}_2$ fusion rule and spin $\Theta = \tfrac{1}{4}$, is not onsiteable. 
More generally, a 1-form $\ZZ_N$ symmetry operator with spin $\Theta = p/(2N)$, where $p$ is odd and $N$ is even, 
cannot be onsiteable.

Indeed, the $\mathbb{Z}_2$ 1-form symmetry associated with the semion exhibits an 't~Hooft anomaly characterized by the 4th cohomology class in $H^4(B^2\mathbb{Z}_2, U(1))$, as shown in Eq.~\eqref{eq: 1/4 B2 B2}. 
Its transgression $\Phi(\omega_4)$ is nontrivial in $H^3(B\mathbb{Z}_2, U(1))$, as given in Eq.~\eqref{eq:1/4 A1 dA1}, and therefore obstructs onsiteability.~\footnote{We note that a semion is onsiteable when realized as an exact $\Z_4$ symmetry, whose $\Z_2$ subgroup is non-faithful and defines \textit{emergent} $\Z_2$ symmetry on ground states. Similarly, a $\Z_n$ 1-form symmetry with spin $1/(2n)$ with even $n$ admits onsite realization as an exact $\Z_{2n}$ symmetry.
See e.g.,~\cite{Ellison2022Pauli} for explicit constructions of lattice models.}


\section{Onsiteability of higher-form symmetries}
\label{sec:higherdim}

In this section, we generalize the previous discussion of 1-form symmetries in $(2+1)$D to higher-form symmetries in higher dimensions.

\subsection{Lattice anomaly of 1-form symmetry in (3+1)D}

For a finite internal 0-form symmetry $G$ in (2+1)D or, more generally, in $(d+1)$D, the vanishing of the 't~Hooft anomaly,
$[\omega_{d+2}] = 0 \in H^{d+2}(BG, U(1))$, 
provides only a necessary condition for onsiteability. 
In general, there exist additional obstructions to onsiteability characterized by a family of indices 
\[
H^{d+2-q}(BG, \mathrm{QCA}_{q-1})~, \qquad \forall~ 0 \le q \le d+1~,
\]
as discussed in Refs.~\cite{shirley2025QCA, tu2025anomaliesglobalsymmetrieslattice}. 
$\mathrm{QCA}_{q-1}$ classifies locality-preserving unitaries in $(q-1)$ spatial dimensions, defined up to multiplication by finite-depth quantum circuits (FDQCs). 
For $q=0$, we define $\mathrm{QCA}_{-1} := U(1)$, recovering the standard group-cohomology anomaly familiar from continuum quantum field theory.

For instance in (2+1)D, there is an index valued in $H^{2}(BG, \mathbb{Q}_+)$, where $\text{QCA}_{1}=\mathbb{Q}_+$ is a multiplicative group of positive rational numbers classifying circuit equivalence of 1d QCAs~\cite{GNVW2012}. Such indices valued in QCAs are sometimes dubbed ``lattice anomalies'', since it obstructs onsiteability and gauging on the lattices though not associated with continuum QFT anomalies.

\subsubsection{Review: $H^2(BG,\mathbb{Q}_+)$ index of 0-form symmetry in (2+1)D}
\label{sec:review lattice anomaly 2d}

We begin by reviewing the definition of the $H^2(BG,\mathbb{Q}_+)$ index for a finite internal 0-form symmetry $G$ in (2+1)D. 
Assume that the $G$ symmetry is generated by a finite-depth circuit $U(g)$ for each $g\in G$, satisfying the multiplication law
\begin{align}
    U(g) U(h) = U(gh)~.
\end{align}
Consider a truncation of these symmetry operators to a disk-shaped region $R$, denoted $U_R(g)$, 
and define a 1d operator $\Omega(g,h)$ supported along the boundary $\partial R$ as
\begin{align}
    \Omega(g,h) := U_R(g) U_R(h) U_R(gh)^{-1}~.
\end{align}
Since $\Omega(g,h)$ preserves locality, one can associate to it a QCA index $\omega_2(g,h)\in \mathrm{QCA}_{1} = \mathbb{Q}_+$. 
The operators $\Omega$ satisfy
\begin{equation}
    \Omega(g,h)\Omega(gh,k) = ({}^g \Omega(h,k))\,\Omega(g,hk)~,
\end{equation}
where ${}^g\Omega$ denotes the conjugation action by $U_R(g)$. 
This relation implies that $\omega_2(g,h)$ is closed, i.e., $\omega_2(g,h)\in Z^2(BG,\mathbb{Q}_+)$. 
Moreover, redefining the truncations $U_R(g)$ can shift $\omega_2$ by a coboundary, 
so the equivalence class $[\omega_2]\in H^2(BG,\mathbb{Q}_+)$ defines a well-defined index.

A nontrivial $H^2$ index signals an obstruction to onsiteability: 
for onsite symmetry operators $U(g)$, one can always choose canonical truncations such that $\Omega(g,h)=1$, 
implying that the corresponding $H^2$ index must be trivial.

\subsubsection{$H^3(B^2G,\mathbb{Q}_+)$ index of 1-form symmetry in (3+1)D}

Here we investigate such ``lattice anomaly'' indices of higher-form symmetry. Let us consider finite 1-form $G$ symmetry in (3+1)D, where we will define an index valued in $H^{3}(B^2G,\mathbb{Q}_+)$ of given symmetry operators.

\begin{figure}[thb]
\centering
\includegraphics[width=0.35\textwidth]{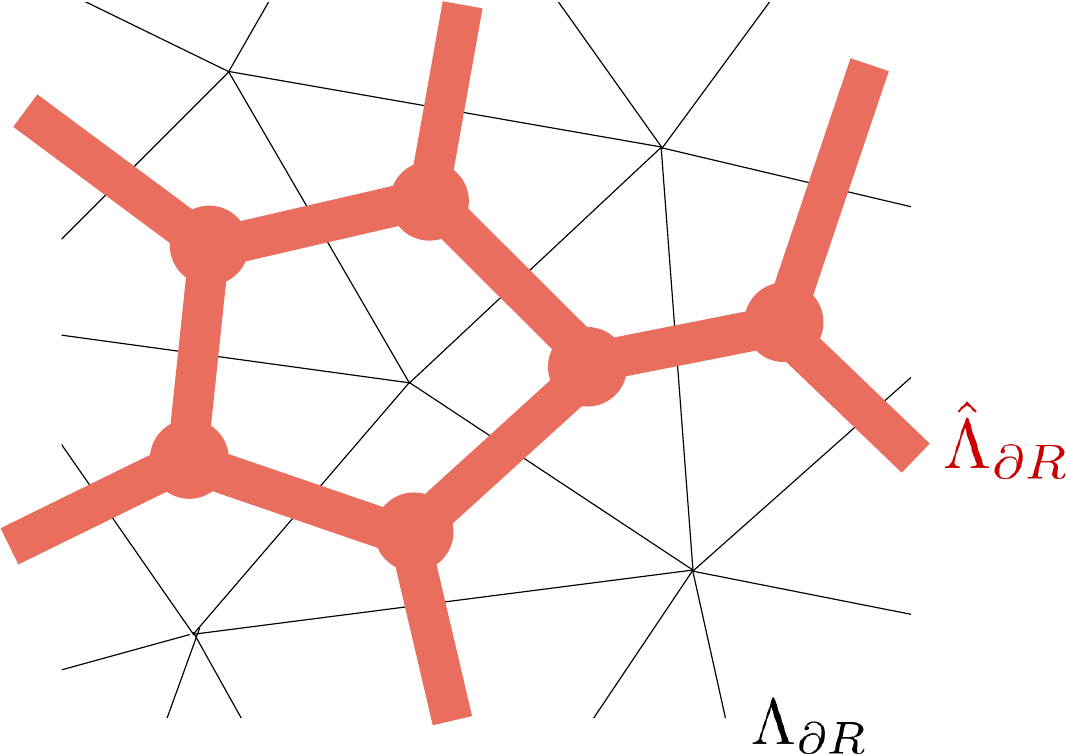}
\caption{For (3+1)D $G$ 1-form symmetry, the operator $\Omega(\epsilon_{01}, \epsilon_{12}, g_{012})$ is a network of 1d QCAs supported at the 2d dual lattice $\hat{\Lambda}_{\partial R}$. Each red edge in the figure carries a 1d QCA, therefore assigns a QCA index valued in $\mathbb{Q}_+$. This assignment defines a 1-cocycle $F\in Z^1(\Lambda_{\partial R}, \mathbb{Q}_+)$.}
\label{fig:QCAnetwork}
\end{figure}

We consider a tensor product Hilbert space in a 3d space, where the 1-form symmetry is defined in the same fashion as Sec.~\ref{sec:reviewof1formanomaly}; the 3d space is endowed with a mesoscopic triangulation $\Lambda$ whose edges are much larger than the locality scale of the lattice model. 
The symmetry operators are finite-depth circuits supported along the 2-cells of the dual lattice (cellulation) $\hat{\Lambda}$. The symmetry operators are again described by Gauss law operators $W^{(g)}_v$, where $v$ is a vertex of $\Lambda$ which corresponds to a single 3-cell of $\hat\Lambda$. $W^{(g)}_v$ is a bubble of a closed surface operator supported along the 2d boundary of this 3-cell. They again satisfy the following three conditions:
\begin{align}
    W_v^{(g)} W_v^{(g')} &= W_v^{(gg')}~, \\
    [\,W_v^{(g)},\, W_{v'}^{(g')}\,] &= 1~, \\
    \prod_v W_v^{(j)} &= 1~, \qquad \forall\, j \in G~. 
\end{align}
Similar to what we have done in Sec.~\ref{sec:reviewof1formanomaly}, the symmetry operator is again labeled by a 0-cochain $\epsilon\in C^0(\Lambda,G)$ by
\begin{equation}
U(\epsilon) = \prod_v \bigl(W_v\bigr)^{\epsilon(v)}\,,
\end{equation}
which satisfies the condition
\begin{equation}
U(\epsilon + dg) = U(\epsilon)~,
\end{equation}
with $dg\in C^0(\Lambda,G)$ a constant (global) cochain with the constant value \(g\in G\). We take a 3d disk region $R$ of $\Lambda$, and consider a truncation of the operator $U(\epsilon)$ within the region $R$, in the form of
\begin{equation}
U_R(\epsilon) = \Bigl(\prod_{v\in \partial R} W_{v;R}^{\epsilon(v)}\Bigr)\;\Bigl(\prod_{v \in \mathrm{Int}(R)} W_p^{\epsilon(v)}\Bigr).
\end{equation}
where $v\in \partial R$ is the vertices $v$ of $\Lambda$ at the 2d boundary $\partial R$. We choose any ordering of operators in the product over $v\in R$.
One then defines a reduced operator  
\begin{equation}
\Omega(\epsilon_{01},\epsilon_{12}, g_{012}) \;=\; U_R(\epsilon_{01})\, U_R(\epsilon_{12})\, U_R(\epsilon_{01} + \epsilon_{12} - dg_{012})^{-1}~.
\label{eq:Omega3d}
\end{equation}
This is a network of 1d line operators supported within edges of the 2d lattice $\hat{\Lambda}_{\partial R}$, which is a dual lattice of a 2d triangulation $\Lambda_{\partial R}$ (the restriction of $\Lambda$ to $\partial R$). See Fig.~\ref{fig:QCAnetwork}.

Each line operator of $\Omega(\epsilon_{01},\epsilon_{12}, g_{012})$ on an edge of $\hat{\Lambda}_{\partial R}$ carries a QCA index valued in $\mathbb{Q_+}$.
This assignment of the QCA index on each dual edge defines a 1-cocycle
\begin{align}
    F(\epsilon_{01},\epsilon_{12}, g_{012})\in Z^1(\Lambda_{\partial R}, \mathbb{Q}_+)~.
\end{align}
Since $\Omega$ follows the equation
\begin{equation}
\Omega(\epsilon_{01},\epsilon_{12},g_{012})\Omega(\epsilon_{02},\epsilon_{23},g_{012})
    =~^{\epsilon_{01}}\Omega(\eps_{12},\eps_{23},g_{123}) \Omega(\eps_{01},\eps_{13},g_{013})~,
\end{equation}
the 1-cocycle $F$ also satisfies 
\begin{equation}
F_e(\epsilon_{01},\epsilon_{12},g_{012})F_e(\epsilon_{02},\epsilon_{23},g_{023})
    =F_e(\eps_{12},\eps_{23},g_{123})F_e(\eps_{01},\eps_{13},g_{013})~.
    \label{eq:2cocycle of F}
\end{equation}
on each edge $e$ of $\Lambda_{\partial R}$.
Because the 2d surface $\partial R$ is simply-connected by construction, the cocycle $F$ is exact, therefore one finds a 0-cochain 
$A(\epsilon_{01}, \epsilon_{12}, g_{012}) \in C^0(\Lambda_{\partial R}, \mathbb{Q}_+)$ such that $F = d A$. For each site $v$, the functional $A_v(\eps_{01},\eps_{12},g_{012})$ depends on the value of $0$-forms $\eps_{01}$ and $\eps_{02}$ at $v$, as well as the group element $g_{012}$. 
Given that $F$ satisfies \eqref{eq:2cocycle of F}, $A$ is also subject to a similar condition used to characterize the anomaly index,
\begin{equation}
\omega_3(g_{012},g_{023},g_{013}, g_{123}) = A_v(\epsilon_{01},\epsilon_{12},g_{012})A_v(\epsilon_{02},\epsilon_{23},g_{023})
    (A_v(\eps_{12},\eps_{23},g_{123})A_v(\eps_{01},\eps_{13},g_{013}))^{-1}~.
\end{equation}
Importantly, this phase $\omega_3$ is a function of group elements $\{g_{ijk}\}$ alone, independent of the 0-cochains $\epsilon_{ij}$. To see this, let us consider an open string (interval) $I$ given by a collection of edges of $\Lambda$. Let us denote the endpoints of $I$ by $v,v'$. By integrating $F=dA$ along the interval $I$, we have $\int_I F =A_{v}A_{v'}^{-1}$. The relation \eqref{eq:2cocycle of F} then implies $\omega_3(v)=\omega_3(v')$. Now, since $A, \omega_3$ are local functionals of 0-cochains $\epsilon_{ij}$ that depend on its value at $v$ alone, $\omega_3(v)=\omega_3(v')$ with $v\neq v'$ implies that $\omega_3(v)$ is independent of $\epsilon_{ij}$.

Let us introduce a simple notation $\omega_3(0123):=\omega_3(g_{012},g_{023},g_{013}, g_{123})$. Then it satisfies
\begin{align}
    \delta\omega_3(01234)= \omega_3(0123)\omega_3(0124)^{-1}\omega_3(0134)\omega_3(0234)^{-1}\omega_3(1234) = 1.
\end{align}
This implies $\omega_3\in Z^3(B^2G,\mathbb{Q}_+)$. Further, there is an ambiguity to redefine $A$ satisfying $F=dA$, shifting by a constant $A(\epsilon_{01},\epsilon_{12},g_{012})\to A(\epsilon_{01},\epsilon_{12},g_{012})\times \chi(g_{012})$ with $\chi(g_{012})\in\mathbb{Q}_+$. This redefines $\omega_3$ by a coboundary $\delta\chi$. Therefore this defines an index valued in $[\omega_3]\in H^3(B^2G, \mathbb{Q}_+)$.

\subsection{Transgression of ``lattice anomaly'' index and onsiteability}

Similar to the 1-form symmetry in (2+1)D discussed in Sec.~\ref{sec:onsiteability 2+1D}, onsiteability is again obstructed by transgression of the index $\omega_3$; $\Phi([\omega_3])\in H^2(BG,\mathbb{Q}_+)$.

The logic is in parallel to Sec.~\ref{sec:onsiteability 2+1D}. 
We consider the 2-cycle $\hat\gamma$ of the dual lattice $\hat\Lambda$ which cuts $R$ into bipartition. The region $R$ is separated into half disks $R_u,R_d$, and $\partial R$ into intervals $(\partial R)_u, (\partial R)_d$. 

For a given constant $x$, we again define a 1-cochain $\iota_*x$ on $\Lambda$ by
\begin{align}
    \iota_*x(v) = 
    \begin{cases}
        x & \text{$v\in \Lambda_u$} \\
        0 & \text{$v\in \Lambda_d$}~.
    \end{cases}
\end{align}
Then, the operator $U(\iota_*h)$ with $h\in G$ is a 2d operator supported along $\hat{\gamma}$. By regarding $U(\iota_*h)$ as a 0-form symmetry operator in 2d along $\hat{\gamma}$, one can define an index $[\omega_2]\in H^2(BG,\mathbb{Q}_+)$ reviewed in 
Sec.~\ref{sec:review lattice anomaly 2d}. By its definition, the index is given by
\begin{align}
    \omega_2(h_{01}, h_{12}) = F_e(\iota_*h_{01}, \iota_*h_{12},0)~,
\end{align}
where $e$ is an edge of $\Lambda$ at $\partial R$ intersecting $\hat{\gamma}$, and $h_{ij}\in G$ such that $\delta h_{ijk}=h_{ij}+ h_{jk} - h_{ik}=0$. The 1-cocycle is written as a coboundary $F=dA$, with a 0-cochain $A$ satisfying
\begin{align}
    A(\iota_*h_{01}, \iota_*h_{12}) = \iota_*\omega_2(h_{01},h_{12})~.
\end{align}
In particular, at a vertex $v$ of $\Lambda_{(\partial R)_u}$, 
\begin{align}
    A_v(\iota_*h_{01}, \iota_*h_{12}) = \omega_2(h_{01},h_{12})~.
    \label{Av=omega2}
\end{align}
Meanwhile, for generic 0-cochains $\epsilon_{ij}\in C^0(\Lambda,G)$,
\begin{align}
    \delta A_v(\epsilon_{01},\epsilon_{12},\epsilon_{23}) = \omega_3~.
\end{align}
Let us write the values $\epsilon_{ij}(v)= h_{ij}\in G$. 
In general $h_{ij}$ is not closed under the coboundary $\delta$;
$g_{ijk} = (\delta h)_{ijk}$. Also, $A_v$ depends on $\epsilon_{ij}$ through their values at $v$, therefore one can write
\begin{align}
    \delta A_v(\{h\}) = \omega_3(\{\delta h\})~.
\end{align}
Since $\omega_3\in Z^3(B^2G,\mathbb{Q}_+)$, $\omega(\delta h)$ is expressed as a coboundary using some 2-cochain $\chi$,
\begin{align}
    \omega_3(\{\delta h\}) = \delta \chi(\{h\})~.
    \label{omega3 = dchi}
\end{align}
This implies that 
\begin{align}
    A_v(\{h\}) = \chi(\{h\})~.
\end{align}
Now let us take $\{h\}$ such that $\delta h=0$. Due to Eq.~\eqref{Av=omega2}, by writing $A_v$ as a functional of $\epsilon_{ij}(v)$ we have
\begin{align}
    A_v(\{h\}) = \omega_2(\{h\}) \quad \text{when $\delta h=0$~.}
\end{align}
By the above two equations, we get
\begin{align}
    \omega_2(\{h\})=\chi(\{h\}) \quad \text{when $\delta h=0$~.}
    \label{omega2 = chi}
\end{align}
Eqs.~\eqref{omega3 = dchi} and \eqref{omega2 = chi} together imply that $\Phi(\omega_3)=\omega_2$ using the transgression map $\Phi$. This cohomology class $\Phi(\omega_3)\in H^2(BG,\mathbb{Q}_+)$ defines an obstruction to onsiteability.~\footnote{An example of non-onsiteable 1-form symmetry in (3+1)D is found in (3+1)D $\Z_2$ toric code with an emergent fermion, which has a $\Z_4$ 1-form symmetry with nontrivial $\omega_2=\Phi(\omega_3)\in H^2(BG,\mathbb{Q}_+)$ (see Sec.~3.2 of Ref.~\cite{Barkeshli2023Codimension}). }

\subsection{Transgression of QFT anomaly}

Let us consider $p$-form $G$ symmetry in generic $(d+1)$ spacetime dimensions. While it is not straightforward to define the ``lattice anomaly'' indices for finite-depth circuits in generic dimensions, we study onsiteability conditions by restricting ourselves to specific structure of the tensor product Hilbert space and symmetry actions.
Within this assumption described later, the ``lattice anomaly'' indices with coefficients in $\text{QCA}_{k}$ are trivial with any $k\ge 0$, i.e., only genuine QFT anomaly $H^{d+2}(B^{p+1}G, U(1))$ is present. We demonstrate that the transgression $\Phi$ of the 't Hooft anomaly index $[\omega]\in H^{d+2}(B^{p+1}G,U(1))$ defines an 't Hooft anomaly of symmetry operators restricted to a codimension-1 submanifold, namely an obstruction to 1-gauging the symmetry. Therefore, the cohomology class obtained by iterative transgressions $\Phi^p([\omega])\in H^{d-p+2}(BG,U(1))$ defines an obstruction to onsiteability.

\subsubsection{Review: general Else-Nayak process for specific symmetry actions}
First, we review the Else-Nayak type reduction for certain types of higher-form symmetry actions. Specifically, suppose that we have a triangulation $\Lambda_M$ of a $d$-dimensional manifold $M$. Assume that the total Hilbert space is given by the tensor product of $|R|$-dimensional local Hilbert spaces on each $p$-simplex of $\Lambda_M$, with $R$ being a finite $G$-module. Therefore, the basis states of the total Hilbert space are labeled by degree $p$ cochains $a_M\in C^p(\Lambda_M,R)$. We further make an assumption on how the $p$-form $G$ symmetry acts on the Hilbert space. A symmetry operator is associated with a $(p-1)$-cochain $\epsilon \in C^{p-1}(\Lambda_M, G)$, and we assume that its action is given by
\begin{equation}
    U_M(\eps)\ket{a_M}=e^{2\pi i\int F_M[a_M,\eps]}\ket{a_M+d\eps}~,\label{UM form}
\end{equation}
where $F_M[a_M,\eps]$ is a local functional of $a_M$ and $\eps$. As described in Ref.~\cite{feng2025higherformanomalies}, this operator generates a $p$-form $G$ symmetry if it satisfies a pair of conditions,
\begin{align}
    U(\epsilon_1+\epsilon_2) = U(\epsilon_1)U(\epsilon_2)~,
\end{align}
and 
\begin{align}
    U(d\chi) =1~, 
\end{align}
for a cochain $\chi\in C^{p-2}(\Lambda_M,G)$. 

To perform the reduction, we consider an abstract triangulated space that is distinct from our physical space $M$. Each $k$-cell in this space with $k\geq 1$ is labeled by a $(p-k)$-form $\eta^{(k)}\in C^{p-k}(\Lambda_M,G)$, and each point is labeled by a different configuration $a_M\in C^p(\Lambda_M,R)$. These labels satisfy the relation 
\begin{equation}
    \delta_M \eta^{(n)}=d\eta^{(n+1)}~,
\end{equation}
where $\delta_M$ is the differential operator in this abstract space. From this construction, $F_M[a_M,\eps]$ can be seen as a functional defined on each labeled $1$-cell that satisfies
\begin{equation}
    dA_{d-1,M}=\delta_M F_M~.\label{dA and F}
\end{equation}
Performing $\delta_M$ on both sides of Eq.~\eqref{dA and F}, we obtain
\begin{equation}
    d(\delta_M A_{d-1,M})=0~,
\end{equation}
and thus there exists a $A_{d-2,M}$ with
\begin{equation}
    dA_{d-2,M}=\delta_M A_{d-1,M}~.
\end{equation}
Repeating this process, we can define functionals $A_{d-k,M}$ on each labeled $(k+1)$-cell that satisfy
\begin{equation}
    dA_{j-1,M}=\delta_MA_{j,M}~,
\end{equation}
and when the degree becomes $(-1)$, we obtain a $U(1)$-valued function $\omega_M=A_{-1,M}$. The function $\omega_M$ only depends on $g\in G$ labels on $(p+1)$-cells, and has the cocycle property
\begin{equation}
    \delta_M\omega_M=0~,
\end{equation}
which makes $[\omega_M]$ a cohomology class in $H^{d+2}(B^{p+1}G,U(1))$. We regard $[\omega_M]$ as the Else-Nayak index of $U_M$. 

\subsubsection{The relation between embedding and transgression}

Now suppose that there is a $(d-1)$-dimensional oriented submanifold $N$ of $M$, whose triangulation $\Lambda_N$ forms a subcomplex of $\Lambda_M$. There is a way of restricting the action of $U_M$ to this submanifold $N$ called the embedding pullback. We refer to Appendix \ref{app:The embedding map of cochains} for details of its definition. 

If we choose a simplicial tubular neighborhood $(\Gamma_N,f_\perp)$ with a sufficiently large radius, then the embedding pullback $U_N$ of $U_M$ will be entirely supported on $\Gamma_N$, and therefore $U_N$ becomes a $(p-1)$-form symmetry on $N$. According to Eq.~\eqref{def embedding pullback}, we have a relation between these higher-form symmetries
\begin{equation}
    U_N(\eta)=U_M(\iota_* \eta)~,\label{UN and UM}
\end{equation}
where $\eta\in C^{p-2}(\Lambda_N,G)$.  Substitute Eq.~\eqref{UN and UM} into Eq.~\eqref{UM form}, we have
\begin{equation}
    U_N(\eta)\ket{a_M}=e^{2\pi i\int F_M[a_M,\iota_*\eta]}\ket{a_M+\iota_*(d\eta)}~.\label{UN form}
\end{equation}
From the locality of $F_M$, we can restrict both sides of Eq.~\eqref{UN form} to the simplicial tubular neighborhood $\Gamma_N$. Suppose that $a_N$ is the restriction of $a_M$ to $\Gamma_N$, then
\begin{equation}
    U_N(\eta)\ket{a_N}=e^{2\pi i\int_{\Gamma_N} F_M[a_N,\iota_*\eta]}\ket{a_N+\iota_*(d\eta)}~.
\end{equation}
Note that $\Gamma_N$ could be decomposed into vertical fibers. The vertical fiber on a simplex $\sigma_{p-1}$ is a $p$-cell $f_\perp(\sigma_{p-1})$ that is homeomorphic to $\sigma_{p-1}\times I$. By grouping $p$-simplices of each vertical fiber together, we can treat $a_N$ as an element in $C^{p-1}(\Lambda_N, S)$, where the $G$-module $S$ is the tensor product of all local Hilbert spaces on the vertical fiber $f_\perp(\sigma_{p-1})$ of $\sigma_{p-1}$. It is easy to see that 
\begin{equation}
    a_N\mapsto a_N+\iota_*(d\eta)
\end{equation}
gives an action of $G$ on $S$. Also, we can integrate the value of $F_M$ on each vertical fiber $f_\perp(\sigma_{d-1})$ together to form a functional $F_N[a_N,\eta]$. Consequently, we obtain 
\begin{equation}
    U_N(\eta)\ket{a_N}=e^{2\pi i\int_N F_N[a_N,\eta]}\ket{a_N+\iota_*(d\eta)}~,
\end{equation}
where
\begin{equation}
    F_N[a_N,\eta]=\pi_*F_M[a_N,\iota_*\eta]~.\label{FN and FM}
\end{equation}
We use $\pi_*$ to denote the integration on vertical fibers. Both $\iota_*$ and $\pi_*$ are linear and commute with the differential operator. 

To obtain the anomaly of the symmetry $U_M$ and $U_N$, we need to perform the generalized Else-Nayak reduction process to the functionals $F_M$ and $F_N$~\cite{feng2025higherformanomalies}. Denote the intermediate steps of reduction by $A_{k,M}$ and $A_{l,N}$, then they satisfy the relations
\begin{equation}
    dA_{k-1,M}=\delta_M A_{k,M}~\label{reduction M}
\end{equation}
and
\begin{equation}
    dA_{l-1,N}=\delta_N A_{l,N}~,\label{reduction N}
\end{equation}
where $\delta_M$ and $\delta_N$ are coboundary operators of labeled simplices, as mentioned in Ref.~\cite{feng2025higherformanomalies}. Given that the expression of Eqs.~\eqref{reduction M} and \eqref{reduction N} involve only linear combinations and the differential operator, a relation similar to Eq.~\eqref{FN and FM} will survive the reduction. Therefore, we obtain from an inductive argument that
\begin{equation}
    A_{k-1,N}[a_N,\{\eta\}]=\pi_* A_{k,M}[a_N,\{\iota_*\eta\}]~,\label{AN and AM}
\end{equation}
in which $\{\eta\}$ stands for the collection of all independent labels in the functional. Specifically, if we choose $k=0$, then left hand side is just the anomaly index $\omega_N$ of $U_N$, which only depends on $(-1)$-form labels $g\in G$. Analogously, the right hand side becomes a $0$-form functional $A_{0,M}$, which only depends on $0$-form and $(-1)$-form labels. Now, Eq.~\eqref{AN and AM} becomes
\begin{equation}
    \omega_N(\{g_N\})=\pi_*A_{0,M}[\{\iota_*g_N\},0]~,
\end{equation}
where $\iota_*g$ is a $0$-form that takes value $g$ above $N$ and $0$ below $N$, and $\pi_* \alpha$ is the difference of the $0$-form $\alpha$ between two sides. Consequently, we obtain
\begin{equation}
    \omega_N(\{g_N\})=A_{0,M}[\{dg_N\},0]~,\label{omegaN and A0M}
\end{equation}
where $dg$ is a $0$-form that takes constant value $g$. 

Finally, we claim that the right hand side of Eq.~\eqref{omegaN and A0M} is exactly the transgression of $\omega_M$. To prove this, we first write down the last reduction step of $F_M$ below
\begin{equation}
    \omega_M(\{g_M\})=\delta A_{0,M}[\{\eta_M\},\{g_M\}]~,\label{last red of M}
\end{equation}
with $0$-form labels $\{\eta_M\}$ and $(-1)$-form labels $\{g_M\}$ that satisfy
\begin{equation}
    dg_M=\delta \eta_M~.\label{gM and etaM}
\end{equation}
Suppose that $\eta_M=dg_N$ for some general $(-1)$-forms $g_N$ that do not necessarily satisfy $\delta g_N=0$, then $g_M=\delta g_N$ and we obtain from Eq.~\eqref{last red of M} that
\begin{equation}
    \omega_M(\{\delta g_N\})=\delta A_{0,M}[\{dg_N\},\{\delta g_N\}]~.\label{omegaM and A0M}
\end{equation}
Now the combination of Eq.~\eqref{omegaM and A0M} and Eq.~\eqref{omegaN and A0M} gives exactly the transgression map, and we obtain that $\omega_N$ is the transgression of $\omega_M$.

\section{Conclusions}

In this paper, we have systematically investigated the conditions under which higher-form symmetries in lattice models can be made onsite, thereby developing a general theory of \textit{onsiteability}. We established that for finite 1-form symmetries in (2+1)D, onsiteability is equivalent to the triviality of the transgression $\Phi([\omega_4]) \in H^3(BG,U(1))$ of the anomaly index $[\omega_4] \in H^4(B^2G,U(1))$, which characterizes the obstruction to ``1-gauging'' the symmetry. This equivalence provides a lattice realization of the correspondence between onsiteability and higher gauging. We further demonstrated that any onsiteable 1-form symmetry in $(2+1)$D can be transformed into transversal Pauli operators by introducing ancillas and finite-depth circuits, thereby showing that onsiteability guarantees a particularly local and transparent operator structure. 

Extending the analysis to (3+1)D and beyond, we introduced the ``lattice anomaly'' indices $[\omega_{d+2-q}] \in H^{d+2-q}(B^{p+1}G,\mathrm{QCA}_{q-1})$, which capture obstructions to onsiteability beyond the continuum ’t~Hooft anomalies. In particular, for 1-form symmetries in (3+1)D, we constructed an explicit ``lattice anomaly'' index $[\omega_3] \in H^3(B^2G,\mathbb{Q}_+)$ whose transgression $\Phi([\omega_3]) \in H^2(BG,\mathbb{Q}_+)$ diagnoses the failure of onsiteability. These results suggest a general correspondence between onsiteability and possibility of higher gauging on lattices: a finite $p$-form $G$ symmetry in $(d+1)$D is onsiteable if and only if all suspended ``lattice anomaly'' indices vanish in cohomology, $\Phi^p([\omega_{d+2-q}]) = 0$ in $H^{d+2-p-q}(BG,\mathrm{QCA}_{q-1})$.

\section*{Acknowledgments}
We thank Andrea Antinucci, Yichul Choi, Kansei Inamura, Ho Tat Lam, Du Pei, Sakura Schafer-Nameki, Shinsei Ryu and Wilbur Shirley for discussions.
Y.-A.C. is supported by the National Natural Science Foundation of China (Grant No.~12474491), and the Fundamental Research Funds for the Central Universities, Peking University.
P-S.H. is supported by Department of Mathematics, King's College London.
R.K. is supported by the U.S. Department of Energy, Office of Science, Office of High Energy Physics under Award Number DE-SC0009988 and by the Sivian Fund at the Institute for Advanced Study. R.K. and P.-S.H. thank the Kavli Institute for Theoretical Physics for hosting the program “Generalized Symmetries in Quantum Field Theory: High Energy Physics, Condensed Matter, and Quantum Gravity” in 2025, during which part of this work was completed. This research was supported in part by grant no. NSF PHY-2309135 to the Kavli Institute for Theoretical Physics (KITP).

\appendix

\section{Review of the Else–Nayak index in (1+1)D lattice systems}
\label{app:elsenayak}
In this appendix, we summarize the microscopic construction of the anomaly index $[\omega]\in H^3(BG,U(1))$ for a 0-form symmetry $G$ acting on a (1+1)D lattice system, following the framework of Ref.~\cite{else2014}.
Consider a $G$ global symmetry operation $U(g)$, where the operators satisfy the group multiplication law $U(g)U(h)=U(gh)$ for $g,h\in G$. We assume that $U(g)$ can be implemented by a finite-depth quantum circuit. To probe the 't Hooft anomaly of the symmetry, restrict its action to an interval $I$ of the chain, and define the partially supported operator $U_I(g)$. Because the global operator is finite-depth, the product $U_I(g)U_I(h)$ differs from $U_I(gh)$ only near the boundaries of $I$. We therefore write
\begin{equation}
U_I(g)U_I(h)=\Gamma_{\partial I}(g,h)U_I(gh)~,
\end{equation}
where $\Gamma_{\partial I}(g,h)$ is a boundary unitary localized around the endpoints of $I$. This yields the associativity condition
\begin{equation}
\Gamma_{\partial I}(g,h)\Gamma_{\partial I}(gh,k)
=(^g\Gamma_{\partial I}(h,k))\Gamma_{\partial I}(g,hk)~.
\end{equation}
Let the two endpoints of $I$ be $l$ and $r$, and decompose the boundary operator as $\Gamma_{\partial I}=\Gamma_l\Gamma_r$.
Focusing on one end, say $l$, the relation above implies
\begin{equation}
\Gamma_{l}(g,h)\Gamma_{l}(gh,k)
=\omega(g,h,k)(^g\Gamma_{l}(h,k))\Gamma_{l}(g,hk)~,
\end{equation}
where $\omega(g,h,k)\in U(1)$ captures the failure of strict associativity.
The collection of phases $\omega(g,h,k)$ satisfies the cocycle condition
\begin{equation}
\omega(h,k,\ell)\omega(g,hk,\ell)^{-1}
\omega(g,h,k\ell)\omega(g,h,k)^{-1}=1~,
\end{equation}
and hence defines an element $\omega\in Z^3(G,U(1))$.
A change of local phase convention $\Gamma_l(g,h)\to \chi(g,h)\Gamma_l(g,h)$ shifts $\omega$ by a coboundary, and therefore the invariant is characterized by the cohomology class
\begin{equation}
[\omega]\in H^3(BG,U(1)).
\end{equation}
This cohomology class, known as the Else–Nayak index, provides a lattice definition of the ’t Hooft anomaly for 0-form global symmetries in (1+1)D systems.

\section{Pauli realization of onsiteable 1-form symmetries in (2+1)D: Duality viewpoint}

When a 1-form symmetry in (2+1)D is onsiteable, we can further transform the symmetry into a transversal Pauli operator by tensoring with ancillas and finite-depth circuits. This has been explicitly shown in Sec.~\ref{subsec:pauli}. In what follows, we offer a field-theoretic intuition for why such an onsite 1-form symmetry can be represented by transversal Pauli operators. The argument below is heuristic and serves mainly to give a physical picture rather than a microscopic proof.

When a quantum system in (2+1)D has $\mathbb{Z}_N$ 1-form symmetry, the anomaly is captured by the topological spin $\frac{p}{2N}$ mod 1 for integer $p$, which corresponds to a T-junction invariant \cite{Levin2003Fermions}. For even $p$, the transgression of the 't Hooft anomaly gives trivial anomaly for (1+1)D 0-form symmetry, while the anomaly is nontrivial for odd $p$ (see Sec.~\ref{subsec:semion}). We will focus on even $p$, where the 1-form symmetry is onsiteable.

We start with a setup where the 1-form symmetry in (2+1)D is made onsite; each Gauss law operator $W^{(g)}_p$ is generated by a product of onsite operators $O^{(g)}_j$ at each onsite Hilbert space $\mathcal{H}_j$. The procedure of Sec.~\ref{subsec:pauli} explicitly transforms this onsite symmetry into a transversal Pauli operator $O'^{(g)}_j$ acting on the ancilla. 
Suppose that the ancilla is $\mathbb{Z}_N$ Pauli stabilizer model which describes the $\Z_N$ gauge theory at low energy (e.g., $\Z_N$ toric code in (2+1)D), and that the Pauli stabilizer model has $\Z_N$ onsite 1-form $G$ symmetry given by transversal Pauli operators $(O'^{(g)}_j)^\dagger$. 
Then, the realization of symmetries in transversal Pauli operators is understood through the duality \cite{Cordova:2017vab}
\begin{equation}
    {\cal S}\quad \longleftrightarrow\quad \frac{{\cal S}\times(\mathbb{Z}_N\text{ TC})}{\mathbb{Z}_N}~,
    \label{eq:duality}
\end{equation}
where $\mathcal{S}$ is the original theory, and $(\mathbb{Z}_N\text{ TC})$ is the ancillary $\Z_N$ Pauli stabilizer model. $()/\Z_N$ represents gauging the diagonal $\Z_N$ symmetry.
The $\Z_N$ 1-form symmetry of $\mathcal{S}$ carries the 't Hooft anomaly $p$, while the Pauli stabilizer model has $\Z_N$ 1-form symmetry with the anomaly $-p$. Therefore the diagonal symmetry is anomaly free and hence gaugeable. By gauging the diagonal symmetry, the diagonal onsite symmetry operator $O^{(g)}_j\otimes (O'^{(g)}_j)^\dagger$ corresponds to the Gauss law operator for gauging the 1-form symmetry, and we have the Gauss law constraint $O^{(g)}_j\otimes (O'^{(g)}_j)^\dagger=1$ by gauge fixing the dynamical 2-form gauge field, where a 2-form gauge fields are regarded as an additional ancillary degrees of freedom. Due to the Gauss law constraint, the symmetry action of the original model $O^{(g)}_j$ is identical to the transversal Pauli operator $O'^{(g)}_j$ of the Pauli stabilizer model. It is expected that gauge fixing is performed by a finite depth circuit acting on the whole Hilbert space, therefore gauging the diagonal 1-form symmetry and gauge fixing yields equivalence of the given 1-form symmetry operators to transversal Pauli operators.

\subsection{Application to bounds on entanglement entropy}

We remark that the onsiteability of anomalous 1-form symmetry also has application for the entanglement entropy of the system following the argument in section 3.4 of \cite{Hsin:2023jqa}, which requires the symmetry operator for entangling region $R$ of ball shape on sphere to factorize into that on $R$ and that on the complement $\bar R$. 
For $\mathbb{Z}_N$ 1-form symmetry generated by line operator of statistics $\frac{p}{2N}$ mod 1 with even $p$ for onsiteability, the projective representation of the truncated symmetry operator $U$ on region $R$ of boundary length $L$ has minimal dimension $d_M=n^{L-1}$ where $n=N/\gcd(N,p)$: they can be described by the auxiliary vector space spanned by $\{\prod_{p=1}^{L-1}U^k_p |0\rangle:k=0,1,\frac{N}{\gcd(N,p)}-1,p=1,\cdots,L-1\}$ for an auxiliary reference vector $|0\rangle$, where $U_p$ for $p=1.\cdots,L-1$ labels different symmetry operators starting from a reference point on the boundary (see Fig.~1 of \cite{Hsin:2023jqa}).
Different vectors in the above basis can be distinguished by eigenvalue under another symmetry operator that braids with $U_p^k$ for different $k,p$. The braiding between two symmetry operators $U^r,U^s$ is $e^{\frac{2\pi i p rs}{N}}$, so only the vectors created by $U^k$ for $k=0,\cdots, \frac{N}{\gcd(N,p)}-1$ can be distinguished in this way. 
 This gives a bound on the entanglement entropy on region $R$:
 \begin{equation}
     S(\rho_R)\geq \log d_M=L\log (N/\gcd(N,p))-\log (N/\gcd(N,p))~.
 \end{equation}
 This generalizes the $\mathbb{Z}_2\times\mathbb{Z}_2$ 1-form symmetry example in \cite{Hsin:2023jqa}.

\section{The embedding map of cochains}\label{app:The embedding map of cochains}
 In this appendix, we construct a map that allows us to embed cochains on a submanifold into a higher-dimensional manifold. This map is used in the main text to derive the onsitability conditions of higher-form symmetry. Within this section, we suppose that $M$ is an oriented $D$-dimensional manifold with a fixed triangulation, and that $N$ is an oriented $(D-1)$-submanifold of $M$. Therefore, the submanifold $N$ locally separate $M$ into two disconnected parts. 

 To describe this geometrical picture in a more rigorous way, we need to introduce the concept of a simplicial tubular neighborhood. Note that the normal bundle of $N$ is 1-dimensional and has two directions. Suppose that the triangulation of $M$ is sufficiently fine, then we could translate $N$ perpendicularly in both directions to obtain submanifolds $N_+$ and $N_-$ of the same triangulation of $M$. Now, the subregion of $M$ between $N_+$ and $N_-$ is homeomorphic to $N\times I$, and is called a tubular neighborhood $\Gamma_N$ of $N$. The radius of $\Gamma_N$ is defined to be the largest number $r$ such that
 \begin{equation}
     \begin{array}{cc}
        r\leq \mathrm{dist} (x,y), & \forall x\in N_+\cup N_-,~y\in N~.
     \end{array}
 \end{equation}
 We require $r\geq 1$ so that $\Gamma_N$ is a genuine neighborhood of $N$. It is separated into two disconnected parts $\Gamma_N^+$ and $\Gamma_N^-$ by $N$. In particular, we exclude simplices in $N$ from $\Gamma_N^{\pm}$. 

 Together with $\Gamma_N$ is a structure $f_+$ that maps each $p$-simplex of $N$ to a $(p+1)$-cell in $\Gamma_N^+$. We define this map inductively as follows. For each point $v$ of $N$, define $f_+(v)$ as a path from $v\in N$ to $v'\in N_+$. Then, for each $p$-simplex $\sigma_p$ of $N$, we define $f_+(\sigma_p)$ as a $(p+1)$-cell such that
 \begin{equation}
     \partial f_+(\sigma_p)=\sigma_p\cup f_+(\partial \sigma_p)\cup \sigma_p'~,\label{boundf+}
 \end{equation}
 where $\sigma_p'$ is a topologically trivial region on $N_+$. Intuitionally, this cell $f_+(\sigma_p)$ looks like a cylinder with bottom $\sigma_p$ and top $\sigma_p'$. We can similarly define a structure $f_-$ that maps each $p$-simplex of $N$ to a $(p+1)$-cell in $\Gamma_N^-$ by changing all $+$ symbols to $-$ in the above construction. Formally, the simplicial tubular neighborhood of $N$ is defined to be the pair $(\Gamma_N, f_{\perp})$, with $f_\perp(\sigma_p)=f_+(\sigma_p)\cup f_-(\sigma_p)$ the vertical fiber on $\sigma_p$. 

 Given a $(p+1)$-form $\alpha\in C^{p+1}(M,G)$, we define $\pi_*\alpha\in C^p(N,G)$ to be its integration along the vertical fibers, that is
 \begin{equation}
     \pi_*\alpha(\sigma_p)=\int_{f_\perp(\sigma_p)}\alpha~.
 \end{equation}
 This map $\pi_*:C^{p+1}(M,G)\to C^p(N,G)$ is a homomorphism. Also, for $\alpha$ supported on the interior of $\Gamma_N$, we have
 \begin{equation}
     d\pi_*(\alpha)=\pi_*d\alpha~,
 \end{equation}
 which is a direct consequence of Eq.~\eqref{boundf+}.

 Next, we describe a one-sided inverse of $\pi_*$ that geometrically embeds cochains of $N$ into $M$. Given a $p$-cochain $a\in C^p(N,G)$, there is an order-preserving embedding $e_*a$ satisfying
 \begin{equation}
\begin{aligned}
    e_*a(\sigma_p)= \left\{ \begin{array}{cl}
 (-1)^p\cdot a(\sigma_p) & \mbox{if  }
 \sigma_p\subseteq N \\ 0 & \mbox{otherwise}
 \end{array}\right. 
\end{aligned}
\end{equation}
 for each $p$-simplex $\sigma_p$ of $M$. However, it does not commute with the differential operator. To resolve this problem, we replace $e_*$ by 
 \begin{equation}
     i_0(a)=d(e_*a)+e_*(da)~,\label{i0def}
 \end{equation}
 which gives a $(p+1)$-cochain on $M$. Performing differential on both sides of Eq.~\eqref{i0def}, we obtain
 \begin{equation}
     di_0(a)=de_*(da)=i_0(da)~,
 \end{equation}
 indicating that $i_0$ commutes with the differential. Also, from our definition of $e_*$ we know that $i_0(a)|_N=0$, and therefore $i_0(a)$ is supported on $\Gamma_N^+\cup \Gamma_N^-$. We decompose $i_0$ into two parts $i_0(a)=i_+(a)+i_-(a)$, with $i_{\pm}(a)$ supported on $\Gamma_N^{\pm}$ respectively. Now, we claim that $i_+$ and $i_-$ both commute with the differential. This follows from 
 \begin{equation}
     di_+(a)-i_+(da)=-(di_-(a)-i_-(da))\label{i+i-}
 \end{equation}
 in which the support of two sides are disjoint. The `practical' embedding map that we use is thus defined to be the injective map
 \begin{equation}
     \iota_*=i_+: C^p(N,G)\to C^{p+1}(M,G)~,
 \end{equation}
 which is a homomorphism that commutes with the differential. 

 To simplify the notation, we introduce $(-1)$-forms to be just elements in $G$. Also, we further assume that $N$ globally separates $M$ into two parts $M_+$ and $M_-$. Consider 0-forms $a$ with $da$ supported in $\Gamma_N$, then $a$ is a constant in the region above $\Gamma_N^+$ and below $\Gamma_N^-$. We assume that $a=a_0^+$ above $\Gamma_N^+$, and $a=a_0^-$ below $\Gamma_N^-$. Now, we define $\pi_*a$ to be a $(-1)$-form 
 \begin{equation}
     \pi_* a=a_0^+-a_0^-\in G~.
 \end{equation}
 Conversely we define $\iota_* g$ for $g\in G$ to be a 0-form satisfying
 \begin{equation}
 \begin{aligned}
    \iota_*g(v)= \left\{ \begin{array}{cl}
    g & \mbox{if  }
    v\in M_+ \\ 0 & \mbox{if } v\in M_-\cup N
    \end{array}\right.~,
 \end{aligned}
 \end{equation}
 for each vertex $v$ in $M$. Under this convention, the commutativity of $\pi_*$ and $\iota_*$ with $d$ is preserved. Also, we can verify directly by definition that 
 \begin{equation}
     \pi_*\cdot \iota_*(a)=a
 \end{equation}
 for all $p\geq -1$ and $a\in C^p(N,G)$. 

 Finally, we define the embedding pullback of higher-form symmetries using the embedding map $\iota_*$. Suppose that there is a $p$-form symmetry acting on $M$, with its action given by the homomorphism
 \begin{equation}
     W:C^{p-1}(M,G)\to \mathcal{U}_M~,
 \end{equation}
 where $\mathcal{U}_M$ denotes the group of unitary operators in $M$. As a higher-form symmetry, it will satisfy the following properties
 \begin{equation}
     W(\eps_1+\eps_2)=W(\eps_1)W(\eps_2)
 \end{equation}
 and
 \begin{equation}
     W(d\eta)=1
 \end{equation}
 for all $\eps_1,\eps_2\in C^{p-1}(M,G)$ and $\eta\in C^{p-2}(M,G)$. Now, the pullback $\iota^* W$ of $W$ is a $(p-1)$-form symmetry acting on $N$, defined by the following equation
 \begin{equation}
     \iota^*W(\alpha)=W(\iota_*\alpha)\label{def embedding pullback}
 \end{equation}
 where $\alpha\in C^{p-2}(N,G)$. We choose a tubular neighborhood $\Gamma_N$ of $N$, the radius of which is large enough that $\iota^*W$ is supported on the interior of $\Gamma_N$. The Hilbert space on $N$ is defined as the tensor product of all local Hilbert spaces in $\Gamma_N$. 
 
\bibliographystyle{utphys}
\bibliography{bibliography}

\end{document}